\documentclass[aps,showpacs,twocolumn,preprintnumbers,prd]{revtex4}

\IfFileExists{srcltx.sty}{\usepackage[active]{srcltx}}

\usepackage{graphicx}
\usepackage{epsfig}
\usepackage{bm}
\def\gtap{\mathrel{ \rlap{\raise 0.511ex \hbox{$>$}}{\lower 0.511ex
   \hbox{$\sim$}}}} \def\ltap{\mathrel{ \rlap{\raise 0.511ex
   \hbox{$<$}}{\lower 0.511ex \hbox{$\sim$}}}} 
\newcommand{\beq}{\begin{equation}}

\newcommand{\eeq}{\end{equation}}
\newcommand{\bea}{\begin{eqnarray}}
\newcommand{\eea}{\end{eqnarray}}

\newcommand{\eV}{\mbox{$ \ \mathrm{eV}$}}

\begin{document}

\preprint{UCLA/08/TEP/03} 

\title{Delayed pulsar kicks from the emission of sterile neutrinos}

\author{Alexander Kusenko}
\affiliation{Department of Physics and Astronomy, University of California, Los
Angeles, CA 90095-1547, USA }
\author{Bhabani Prasad Mandal} 
\author{Alok Mukherjee}
\affiliation{Department of Physics, Banaras Hindu University,
Varanasi-221005, India}


\begin{abstract}
The observed  velocities of pulsars suggest the possibility that sterile
neutrinos with mass of several keV are emitted from a cooling neutron star.  
The same sterile neutrinos could constitute all or part of cosmological dark
matter.  The neutrino-driven kicks can exhibit delays depending on the mass and the mixing angle,
which can be compared with the pulsar data.  We discuss the allowed ranges of sterile neutrino
parameters, consistent with the latest cosmological and X-ray bounds, which can explain the
pulsar kicks for different delay times. 

\end{abstract}

\pacs{PACS numbers: 97.60.Gb, 14.60.Pq, 97.60.Bw}


\maketitle

Observed velocities of pulsars~\cite{astro} can be explained by an anisotropic
emission of sterile neutrinos~\cite{ks97,fkmp,pulsars,Kusenko:2004mm} or other light
particles~\cite{Farzan:2005yp}.   Sterile neutrinos are firmly rooted in
particle physics~\cite{sterile_neutrino_review}, because the gauge singlet
(right-handed) neutrinos are needed to account for the observed neutrino masses
in what is called the seesaw Lagrangian~\cite{seesaw}.  If some of the gauge
singlets turn out to be light, they can appear below the electroweak scale as
sterile neutrinos.  There are further hints in favor of this intriguing
possibility: in addition to explaining the pulsar kicks, sterile neutrinos with
the same parameters could make up the cosmological dark
matter~\cite{dw,st,Kusenko:2006rh,Fuller,Palazzo:2007gz} and
could play an important role in the formation of the first stars~\cite{reion}.  This form
of ``warm'' dark matter may be in good agreements with observational inferences regarding the
small-scale structure~\cite{cdm-wdm,dSphs}.  In contrast, the
active neutrino oscillations cannot explain the pulsar kicks, unless they have very large magnetic
moments~\cite{voloshin}, or the mass difference is large enough to allow the
Mikheev-Smirnov-Wolfenstein~\cite{msw}  resonance at density
$10^{11}-10^{12}$~g/cm$^3$~\cite{Kusenko:1996sr,eq}, which is excluded by the current data on
neutrino masses. 

The most promising way to discover sterile neutrinos is by observing X-ray
photons from decays of the relic sterile neutrinos, which have lifetimes longer
than the age of the universe but which can, nevertheless, produce a detectable
signal~\cite{x-rays}.  The X-rays produced by the dark matter decays, and the
production rate of sterile neutrinos in a supernova are both governed by two
parameters: the sterile neutrino mass and the mixing angle.   It is, therefore,
important to understand the allowed ranges of these parameters.  The range
implied by the pulsar kicks can help focus the X-ray searches.  In
Ref.~\cite{fkmp}, the allowed range was discussed for both the resonance
and the off-resonance production.   In this paper we will reconsider this range of parameters,
apply the present constraints from the X-ray observations and the Lyman-alpha
bounds, and we will also relate the range of sterile neutrino masses and mixing
angles to the delays in the onset of the kick.   This delay can have some
observable consequences and can be determined from the studies of the pulsar
populations~\cite{Ng:2007aw}.  

In applying the cosmological constraints, we distinguish between two different issues: the
particle's existence and its ability to account for all of dark matter.  Assuming the
standard cosmology (rather than, e.g., the low-reheat cosmology~\cite{low-reheat}), one expects the
neutrino oscillations to generate some out-of-equilibrium population of dark-matter particles that
depends only on the mass and the mixing parameters.  This population may constitute only a fraction
of dark matter if the mixing angle is small enough.  However, there are other ways in which the
relic sterile neutrinos could be produced: for example, they can be produced from the inflaton 
decay~\cite{st} or from the decay of the Higgs boson at the electroweak
scale~\cite{Kusenko:2006rh}.  The latter scenario produces a population of dark matter particles
that is considerably ``colder''~\cite{Kusenko:2006rh} than the warm dark matter originating from
neutrino oscillations, and the amount of dark matter is completely independent from the mixing
angle.   Nevertheless, regardless of any additional production mechanisms, the production by
oscillations cannot be ``turned off'' (except for non-standard cosmological
scenarios~\cite{low-reheat}). Therefore, there exists a robust cosmological bound on the mass and 
mixing angle, which is based on the effects of sterile neutrinos
produced by oscillations, even if they do not make up all the dark matter. 

We, therefore, show two exclusion regions in Fig.~1.  In the solid ``excluded region'', the
existence of a sterile neutrino conflicts with the assumptions of standard cosmology.  Below this
region, but above the dashed line, the particle may exist, but it may not account for all the dark
matter because of the existsing X-ray bounds~\cite{x-rays}.  Finally, below the dashed line, the
particle may exist and may account for all dark matter.   As one can see from the figure, the
resonant mechanism is inconsistent with all dark matter being sterile neutrinos, but is a 
viable explanation for the pulsar kicks, as long as the sterile neutrinos make up only a part of
dark matter. In contrast, the off-resonant mechanism is consistent with all of dark matter being in
the form of sterile neutrinos. 

In a supernova, the sterile neutrinos are produced in two ways: the active
neutrinos can oscillate into the sterile neutrinos on resonance~\cite{ks97},
or off-resonance~\cite{fkmp}.  In both cases, they escape anisotropically
because the electrons and other fermions in the newly formed neutron star are
polarized in the magnetic field.  Of course, the ordinary neutrinos are
produced with some anisotropy as well, but their production asymmetry is
completely washed out by the numerous scatterings the neutrinos undergo on
their way out of the star as they diffuse in approximate thermal
equilibrium~\cite{eq}.  In contrast, the sterile neutrinos escape without
scatterings, with the emission asymmetry equal their production asymmetry,
because their scattering cross section is suppressed by the small mixing
angle. 

In the case of the production off resonance, the allowed range of parameters
has a direct connection with the time delay from the supernova collapse until the onset of the
kick. In the case of production on resonance, the connection is less obvious. While there are many
uncertainties in the supernova parameters, we think it is of interest to show the parametric
dependence of the allowed kick parameters on time delay.  

Let us briefly summarize the results of Ref.~\cite{fkmp}.
The off-resonance production rate of sterile neutrinos is
determined by the mixing angle in matter $\theta_m$, which, in general is not
the same as the mixing angle $\theta$ in vacuum:
\begin{equation}
\sin^2 2 \theta_m = 
\frac{(\Delta m^2 / 2p)^2 \sin^2 2 \theta}{(\Delta m^2 / 2p)^2 \sin^2 
2 \theta + ( \Delta m^2 / 2p \cos 2 \theta - V_m)^2}, 
\label{sin2theta}
\end{equation}
where the matter potential $V_m$ is positive (negative) for $\nu
(\bar{\nu})$, respectively; $p$ is the momentum.  For the case of $\nu_e$
oscillations into sterile neutrinos, 
\begin{equation}
V_m= \frac{G_{\!\!_F} \rho}{\sqrt{2} m_n} (3 Y_e-1+4 Y_{\nu_e}+2Y_{\nu_\mu}
+2Y_{\nu_\tau}). 
\end{equation}
In a core collapse supernova, the initial value of this matter potential is 
$V_m \simeq (-0.2 ...+ 0.5) V_0$, where $V_0$ depends on the density $\rho
$; $V_0= G_{\!\!_F} \rho / \sqrt{2} m_n \simeq 3.8 \eV (\rho / 10^{14}
\mathrm{g cm^{-3}}) $.

 It was pointed out in Ref.~\cite{Fuller} that, in the presence of sterile
neutrinos, rapid conversions between different neutrino
flavors can drive the effective potential to its stable equilibrium fixed
point 
\begin{equation}
V_m \rightarrow 0.  
\end{equation} 
This equilibration takes place on a time scale 
\begin{eqnarray}
\label{timeeqoffres}
 \tau_{_V}  & \simeq &
\frac{4 \sqrt{2} \pi^2 m_n}{G_{\!\!_F}^3 \rho}
\frac{ (V_m^{(0)})^3}{(\Delta m^2)^2 \sin^2 2 \theta } \frac{1}{\mu^3}
\\
& \sim & 
 \frac{10^{-9} {\rm s}}{\sin^2  \theta} 
\left (\frac{V_m^{(0)}}{0.1\, \mathrm{eV}} \right )^3 
\left (\frac{50 \, \mathrm{MeV}}{\mu} \right )^3 \left ( \frac{ 
10 \, \mathrm{keV}^2
}{\Delta m^2
} \right )^2. \nonumber   
\label{timeeqoffresnumerical}
\end{eqnarray}

Once the equilibrium is achieved at $V_m
\approx 0,  $  the effective mixing angle in matter is close to that in
vacuum, and from this point on the emission of sterile neutrinos proceeds at a
much higher rate.   

\begin{figure}
\epsfxsize=8.5cm 
\epsfbox{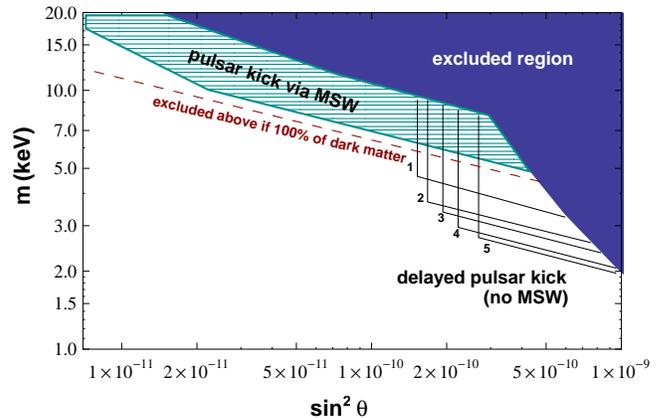}
\caption{ The allowed regions for delayed kicks with delays from 1 through 5 seconds (assuming the
 other parameters are fixed) are shown by
black solid lines marked by the numbers representing the delay time in seconds.  The exclusion
region is based on the combination of the X-ray bounds and the Lyman-$\alpha$ bounds, according to
Palazzo et al.~\cite{Palazzo:2007gz}.  Here we distinguish between the two possibilities: (i)
sterile neutrino with a given mass and mixing angle may exist, and (ii) sterile neutrino with a
given mass and mixing angle may constitute the entire cosmological dark matter.  The former
possibility is viable for all the points below the solid ``excluded region'', while the latter is
limited from above by the dashed line (see discussion in the text).   These bounds can be evaded in
low-reheat cosmologies~\cite{low-reheat}. 
} 
\label{figure:parameters}
\end{figure}

There is a considerable uncertainty in the equilibration time given by 
eq.~(\ref{timeeqoffresnumerical}) because several parameters are functions of
time and position in the star and are not known precisely.  The equilibration
does not have to occur simultaneously in the entire star.   However, although
the emission of both active and sterile neutrinos is subject to much
uncertainty, the total energy must be close to the initial gravitational energy
of the collapsing core, $3\times 10^{53}$~erg.  It is also known that at least
30\% of this energy must be carried out of the supernova by ordinary neutrinos
to explain the observed neutrino signal from SN1987A.  Finally, if the sterile
neutrinos are to explain the pulsar kicks, they must carry a non-negligible
fraction ${\cal  E_{\rm s}} $ of the total energy ${\cal E_{\rm tot}}
$\cite{fkmp}: 
\begin{equation}
r_{\cal E}= 
\left ( \frac{\cal  E_{\rm s}}{\cal E_{\rm tot}} \right ) \approx 0.25.
\label{rE}
\end{equation}
This results in a 1\% anisotropy of the overall neutrino emission and give the
pulsar a kick consistent with observations. 

The relative rates of the active and sterile neutrino production depend on
the mixing angle $\theta$ and the temperatures in the core (where the sterile
neutrinos are produced) and the neutrinosphere (from where the active
neutrinos are emitted).  In addition, the time interval over which the sterile
neutrinos are emitted can be shortened if the equilibration of $V_m\rightarrow
0$ takes some non-negligible time.  Based on the discussion of
Ref.~\cite{fkmp}, one can estimate the ratio $\left ( {\cal  E_{\rm s}} / 
{\cal E_{\rm tot}} \right )= {\cal  E_{\rm s}} / 
({\cal E_{\rm s}}+{\cal E_{\nu }}) $ as follows:
\begin{equation}
\left ( \frac{\cal  E_{\rm s}}{\cal E_{\nu}} \right ) \sim  
\sin^2 \theta \, 
\left ( \frac{T_{\rm core} }{T_{\nu-{\rm
      sphere}}} 
  \right )^6 \left(\frac{t- \tau_{_V}}{t} \right)
 \, f_{_{\rm M}}\,  f_{\rm d.o.f.},  
\label{rE_sin2theta}
\end{equation}
where $f_{_{\rm M}}< 1$ is the fraction of enclosed mass of the core from
which the emission of sterile neutrinos is efficient, and $ f_{\rm d.o.f.}
\ge 1$ is an enhancement due to a possible increase in the effective
degrees of freedom at high density~\cite{prakash}.   

Using eqs.~(\ref{rE}) and (\ref{rE_sin2theta}) one can find the minimal mixing
angle consistent with $r_{\cal E}\ge 0.25$.  The minimal allowed value of the mixing angle
corresponds to the maximal allowed value of the ratio of these two temperatures, 
$T_{\rm core} /{T_{\nu-{\rm sphere}}}$.  The neutrinosphere temperature $ T_{\nu-{\rm
sphere}}=2-5$~MeV~\cite{suzuki,prakash,Prakash:1996xs} is determined by the heat exchange due to
neutrino cooling near the surface of last scattering. This temperature has very little dependence on
the nuclear equation of state and is almost entirely determined by the conditions around the
neutrinosphere, {\it i.e.} at density of the order of $10^{11}-10^{12} \ {\rm g}/{\rm
cm}^3$~\cite{suzuki,prakash,Prakash:1996xs}.  In contrast, the core temperature depends on the
nuclear equation of state at densities $\rho \gg 10^{14}{\rm g}/{\rm cm}^3$ and can vary
dramatically, depending on the assumptions about nuclear matter~\cite{prakash,Prakash:1996xs}.  With
a few exceptions, the models listed in Tables of Ref.~\cite{Prakash:1996xs} predict  the core
temperatures $T_{\rm core} < 100$~MeV, and the majority of these models show the temperatures in the
range $T_{\rm core} \approx (20-70)\, {\rm MeV}$~\cite{Prakash:1996xs}. We will adopt the latter as
the allowed range of the core temperatures. 

Since there is little or no correlation between the core temperature and the neutrinosphere
temperature, we allow the ratio $T_{\rm core} /{T_{\nu-{\rm sphere}}}$ vary between the lowest and
the highest value for  $ T_{\nu-{\rm sphere}}=2-5$~MeV and $T_{\rm core} \approx (20-70)\, {\rm
MeV}$.  The large uncertainty in the ratio of temperatures is the reason for the broad allowed range
of masses and mixing angles.   The corresponding contours are shown in Fig.~1 for different values
of the time delay. The longer time delays correspond to lower mass for the same mixing angle.
However, since the time available for the kick is shorter, the minimal mixing angle increases for
longer delays (see Fig.~1). 

The neutrino-driven kicks have several predictions, in addition to time delays, that can be tested
using the astronomical observations.   The neutrino kick mechanism does not predict a
correlation between the magnitude of the surface magnetic fields and the pulsar velocity (the
``$B-v$'' correlation)~\cite{Kusenko:2004mm}.  The kick velocity is determined by the magnetic field
inside the hot neutron star during the  first seconds after the supernova collapse.  In contrast,
the astronomical observations can be used to infer the surface magnetic fields of pulsars some 
millions of years later.  The relation between the two is highly non-trivial because of the
complex evolution the magnetic field undergoes in a cooling 
neutron star.   However, the correlation of the direction of the spin axis and the direction of the
pulsar velocity is a generic prediction of this mechanism~\cite{sp,Kusenko:2004mm}.  Such a
correlation is confirmed by recent observations~\cite{Omega_corr}.  In the event of a nearby
supernova, the neutrino kick can produce gravity waves that could be detected by LIGO and
LISA~\cite{loveridge,cuesta}.  Finally, the neutrino-driven kicks can increase the energy of the
supernova explosion because they enhance the convection in front of the moving neutron star and
increase the energy of the shock wave~\cite{Fryer:2005sz}, and also because they deposit entropy
ahead of the shock~\cite{Hidaka:2007se}.   The increase of convection in front of the moving neutron
star can produce asymmetric jets with the stronger jet pointing in the direction of the pulsar
motion, in contrast with what one could expect from other kick mechanisms~\cite{Fryer:2005sz}.

The work of A.K. was supported in part by DOE grant DE-FG03-91ER40662 and by
the NASA grant ATFP07-0088.  A.K. appreciates hospitality of the Aspen
Center for Physics.

\end{document}